# Tandem Queueing Systems Maximum Throughput Problem

**Daniel Marian Merezeanu**, "Politehnica" University of Bucharest
**Faculty of Automation and Computers**
Spl. Independentei 313, 77206, Bucharest, Romania
e-mail: daniel@aii.pub.ro

**Daniela Andone**, "Politehnica" University of Bucharest
**Faculty of Automation and Computers**
Spl. Independentei 313, 77206, Bucharest, Romania
e-mail: dana@aii.pub.ro

**Abstract**

In this paper we consider the problem of maximum throughput for tandem queueing system. We modeled this system as a *Quasi-Birth-Death* process. In order to do this we named *level* the number of customers waiting in the first buffer (including the customer in service) and we called *phase* the state of the remining servers. Using this model we studied the problem of maximum throughput of the system: the maximum arrival rate that a given system could support before becoming saturated, or unstable. We considered different particular cases of such systems, which were obtained by modifying the capacity of the intermediary buffers, the arrival rate and the service rates.
The results of the simulations are presented in our paper and can be summed up in the following conclusions:
1. The analytic formula for the maximum throughput of the system tends to become rather complicated when the number of servers increase
2. The maximum throughput of the system converges as the number of servers increases
3. The homogeneous case reveals an interesting characteristic: if we reverse the order of the servers, maximum thruoughput of the system remains unchanged
The QBD process used for the case of Poisson arrivals can be extended to model more general arrival processes.

## INTRODUCTION

In this paper we analyze the problem of maximum throughput of a tandem queueing system. We consider a system with Poisson arrivals and independent exponential service times. This kind of system is appropriate for representing the production lines – the basis of flexible manufacturing systems.
Basically a production line consists of a one–way serial configuration of machines and buffers. To model this as a tandem queues system we will assimilate the machines to servers and the buffers to queues placed between these servers. The tandem queues system obtained is presented in figure 1, This system consists of *K+1* servers (notated $S_i$, where *i=0,1,...,K*) and the same number of buffers (the buffers capacity is notated $B_i$, where *i=0,1,...,K*).

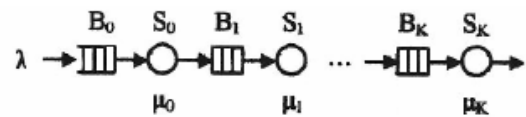

Figure 1 – Tandem queueing system's structure

In order to analyze the problem of maximum throughput for this system we made the following assumptions:
- The first buffer ($B_0$) has as infinite capacity and the intermediary buffers are finite and have the same capacity (we noted this *B*)
- Customers' arrivals into the system (first buffer) are Poisson arrivals
- The service order for each server is first-in-first-out and with independent exponential service times (noted $\mu_0,...\mu_k$, where *i=0,1,...,K* – index of servers $S_i$ which are included in the system) and the time required to service a customer at a particular station is independent of the time required to service that customer at any other station
- Each customer arriving into the system requires service by each server, sequentially
- When server $S_i$, i<K completes service, if server $S_{i+1}$ is occupied and the buffer before it $B_{i+1}$ is full, then server $S_i$ is idled until space becomes available at server $S_{i+1}$ that is, server $S_i$ becomes blocked.

The goal of this paperis to study the maximum arrival rate ($\lambda_{max}$) that the system can support before becoming saturated or unstable, the so-called *maximum throughput*. In order to do this we model as a quasi-birth-and-dead (QBD) process.We simulated the behavior of the QBD process in order to obtain the formula and values for the maximum throughput for different configurations. We present the results of these simulations in the paper.

## GENERATING THE QBD MODEL

In order to model the given tandem queueing system as a QBD process we proceeded as follows. We named *level* the number of customers in the buffer before server $S_0$ (including the customer in service) and we called *phase* the state of the remaining servers. Considering these notations, the only events that can change the state of the system are the arrival of the new customer or a service completion by one of the servers. Since the probability of two of these events occurring simultaneously is zero and the level can icrease or decrease by at most one on the occurrence of a single event, this means we have infinite, continous-time QBD process.

To specify the phase we need only to consider the case in which server $S_0$ is occupied, so we encode the phase as an ordered K-tuple ($m_1,...,m_k$), where $0 \leq m_i \leq B_i + 2$, this means that there are $m_i$ customers in the buffer before server $S_i$, including the one being served. When $m_i = B_i + 2$, this means that there are $B_i + 1$ customers at the server $S_i$ and server $S_i$ is blocking server $S_{i-1}$. It is obvious that a phase with $m_i = 0$ and $m_{i+1} = B_i + 2$ is not a valid one, since it is not possible for the server $S_i$ to be both empty and blocked simultaneously. This allows us to determine the number of valid phases $M$ as:

$$M + \frac{[(B+3)+\sqrt{(B+1)(B+5)}]^{K+1} - [(B+3)+\sqrt{(B+1)(B+5)}]^{K+1}}{2^{K+1}\sqrt{(B+1)(B+5)}}$$

(1)

where B represents the capacity of the intermediary buffers.
We used this model to study the maximum tgroughput of the considered system.

## MAXIMUM THROUGHPUT PROBLEM

In order to study the maximum throughput problem, we should generate the infinitesimal generator matrix Q of the defined QBD process. The Q matrix is a tri-diagonal block matrix having the structure presented below:

$$Q = \begin{bmatrix} B_1 & A_0 & 0 & 0 & \cdots \\ A_2 & A_1 & A_0 & 0 & \cdots \\ 0 & A_2 & A_1 & A_0 & \cdots \\ 0 & 0 & A_2 & A_1 & \cdots \\ \cdots & \cdots & \cdots & \cdots & \cdots \end{bmatrix}$$

(2)

where block $A_1$ accounts for those transitions where the level remains unchanged, $A_0$ accounts for those transitions where the level increases, and $A_2$ accounts for those transitions where the level desreases. Block $B_1$ corresponds to the transitions from level 1 to level 0 (there are no customers in the input buffer $B_0$) and it is of no interest for our purposes.

Having the Q matrix, the next step is to determine tge condition of stability of the system. Let $A - A_0 + A_1 + A_2$ and let $\pi$ be the *M*-dimensional row vector satisfying:

$$\pi A = 0, \ \pi e = 1 \tag{3}$$

where e is a column vector of ones.
Then the system is stable if and only if:

$$\pi A_2 e > \pi A_0 e \tag{4}$$

In computing $A_i$, i=0,1,2 it is enough to consider the case where the level is greater than 0. Then any arrival will cause the level to increase, but the phase will remain the same. In consequence $A_0 = \lambda I$. Therefore, when we compute *A* we will always have an arrival added to the diagonal of A from $A_0$ but by the same token we will always have an arrival subtracted from the diagonal of *A* because of the $-\lambda$ on the diagonal of $A_1$ that we need to make the sum of the rows of *Q* equal to 0. That is, in computing the A matrix, arrivals make no difference and can be ignored.
To get the maximum value of $\lambda$ (maximum throughput) we need to solve:

$$\pi A_2 e = \pi A_0 e \tag{5}$$

Since $A_0 = \lambda I$, equation (5) can be simplified to:

$$\lambda = \pi A_2 e \tag{6}$$

and the problem is reduced to solving the systems:

$$\pi A = 0, \tag{7}$$

Since any equation in the first system is redundant, we simply replace one of them with $\pi e = I$. In order to obtain the maximum throughput ($\lambda_{max}$) for the steady state of the QBD process we only need to solve the resulting system. The only remaining problem is to determine the structure of the blocks for matrix *Q*. The $A_0$ block (corresponding to the transitions to the next level) is easy to obtain. As we already saw, the level can only increase with the arrival of a new customer into the system. This doesn't change the phase, so $A_0$ will be an *M*-diagonal matrix of $\lambda$. Yet the process of generating the other two blocks $A_1$ and $A_2$ is not so straightforward. In order to generate $A_1$ and $A_2$ we should analyze each phase and decide which are the possible transitions and determine the resulting phase and level. We made the observation that the level decreases only at the moment when server $S_0$ completes the service of a customer and may pass it to server $S_1$. There are two possible situations: if server $S_0$ was no the server that completed the service, then the only way th elevel can decrease is if the server $S_1$ is blocking $S_1$. The complexity of the model made this analysis possible only by using numerical methods, so we develop an algorithm to do that. We use this algorithm to simulate different configurations of the system. The following sections describe the simulation algorithm and the results we obtained.

**SIMULATION ALGORITHM**

The algorithm's steps are:
1. Initialize the process; read the number of servers into the tandem queueing system, their rates and the capacities of the intermediary buffers.
2. Determine the number of valid phases and which are these valid phases.
3. Analyze each valid phase in order to find the events that can occur next and generate the blocks of the $Q$ matrix, according to the following rules:
   - for the arrival of a customer, the level increase and the phase remain unchanged; in consequence the block $A_0$ is a diagonal matrix of $\lambda$
   - for a server $S_i$ ending the serving process of a customer (only servers that are not blocked can do this), if the level is not decreasing enter $\mu_i$ into the corresponding position of block $A_1$; otherwise enter $\mu_i$ in the corresponding position of block $A_2$; the diagonal elements of the block $A_1$ are determined by the equalizing with zero the sum of the rows of the $Q$ matrix.
4. Determine the maximum throughput of the system.

We will illustrate this process for a particular configuration, consisting of two servers (*K*-1) and an intermediary buffer with capacity *B*=2. This particular structure is presented in figure 2.

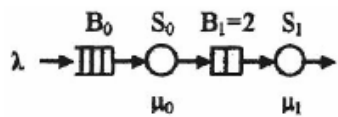

Figure 2 – Example of a tandem queueing system with two servers and a finite intermediary buffer

The associate QBD model for the above tandem queueing is presented in figure 3.

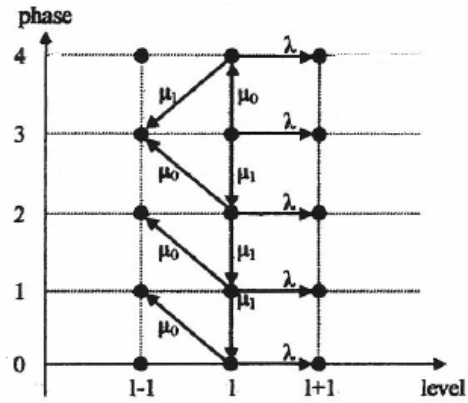

Figure 3 – The associated QBD model of the tandem queueing system

The corresponding blocks $A_0$, $A_1$ and $A_2$ for the above QBD model are:

$$A_0 = \begin{bmatrix} \lambda & 0 & 0 & 0 & 0 \\ 0 & \lambda & 0 & 0 & 0 \\ 0 & 0 & \lambda & 0 & 0 \\ 0 & 0 & 0 & \lambda & 0 \\ 0 & 0 & 0 & 0 & \lambda \end{bmatrix}$$

$$A_1 = \begin{bmatrix} -\Delta_{11} & 0 & 0 & 0 & 0 \\ \mu_1 & -\Delta_{22} & 0 & 0 & 0 \\ 0 & \mu_1 & -\Delta_{33} & 0 & 0 \\ 0 & 0 & \mu_1 & -\Delta_{44} & \mu_0 \\ 0 & 0 & 0 & 0 & -\Delta_{55} \end{bmatrix}$$

$$A_2 = \begin{bmatrix} 0 & \mu_0 & 0 & 0 & 0 \\ 0 & 0 & \mu_0 & 0 & 0 \\ 0 & 0 & 0 & \mu_0 & 0 \\ 0 & 0 & 0 & 0 & 0 \\ 0 & 0 & 0 & \mu_1 & \end{bmatrix} \quad (8)$$

where:
$\Delta_{11} = \mu_0 + \lambda$
$\Delta_{22} = \Delta_{33} = \Delta_{44} = \mu_0 + \mu_1 + \lambda$
$\Delta_{55} = \mu_1 + \lambda$

and the formula of the maximum throughput obtained for this particular configuration is:

$$\lambda_{max} = \frac{\mu_0 \mu_1 \left( \mu_0^3 + \mu_0^2 \mu_1 + \mu_0 \mu_1^2 + \mu_1^3 \right)}{\mu_0^4 + \mu_0^3 \mu_1 + \mu_0 \mu_1^3 + \mu_1^4} \quad (9)$$

**EXPERIMENTAL RESULTS**

We use this simulation algorithm for different configurations of the tandem queueing system having a different number of servers, with different service rates, with or without intermediary buffers. Table 1 summarizes some of the results of the simulation for configuration without intermediary buffers. We change only the service rate of the first server $S_0$ and use the same service rate $\mu_i = 1$ for the others.

Table 1 – Maximum throughput for the configuration without intermediary buffers

| Number of servers | Value of $\mu_0$ | | |
|---|---|---|---|
| | 0.8 | 1.0 | 1.25 |
| 3 | 0.519989942 | 0.564102564 | 0.598437788 |
| 4 | 0.485029352 | 0.514775489 | 0.535049700 |
| 5 | 0.463993704 | 0.485798122 | 0.499168087 |
| 6 | 0.449869861 | 0.466713263 | 0.476185018 |

In figure 4 we present a graphic representation of the maximum throughput of the system.

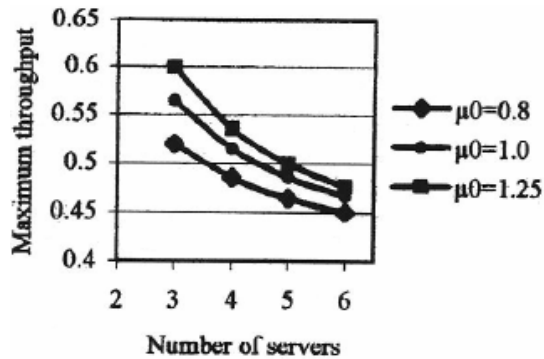

Figure 4 – Maximum throughput for the system without intermediary buffers

The results for the same configurations of the tandem queueing system, using intermediary buffers with capacity B=1 are presented in table 2.

Table 2 – Maximum throughput for the system with one intermediary buffer with capacity B=1 per server

| Number of servers | Value of $\mu_0$ | | |
|---|---|---|---|
| | 0.8 | 1.0 | 1.25 |
| 3 | 0.615528799 | 0.670466159 | 0.707254387 |
| 4 | 0.592393780 | 0.631152686 | 0.652598317 |
| 5 | 0.578207816 | 0.607583286 | 0.621585610 |
| 6 | 0.568521082 | 0.591825779 | 0.601677388 |

The graphic representation of the evolution of maximum throughput for a tandem queueing system with intermediary buffers is shown in figure 5.

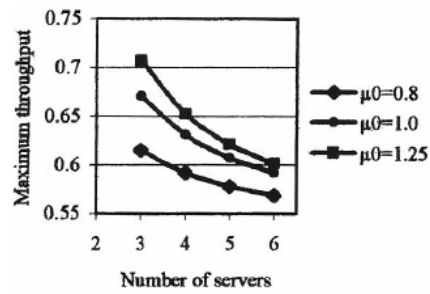

Figure 5 – Maximum throughput for the system with intermediary buffers

## CONCLUSIONS

The results obtained allow us to formulate some interesting conclusions. First we should notice that the complexity of the system significantly increases when the number of servers and the capacity of the analytic formula of the system's maximum throughput tends to become rather complicated.
We also noticed that the evolution of the maximum throughput of a tandem queueing system converges as the number of servers increases. Also the results prove that the intermediary buffers increase the value of the maximum throughput of the system.
Another observation is that the homogeneous structure (all the intermediary buffers having the same capacity) reveals an interesting characteristic, that if the order of the servers is reversed, the maximum throughput of the system remains unchanged. In further research our efforts will focus on extending the application of the QBD process for more general arrival processes.
Our immediate plans are to adapt the method presented in the paper in studying the maximum throughput problem for a tandem queueing system with $MMPP_n$ arrivals and to find a way to improve the simulation algorithm and to reduce its complexity.